\documentclass[twocolumn,pra,showpacs,superscriptaddress]{revtex4}
\usepackage{amssymb}
\usepackage{amsmath}
\usepackage{graphicx}
\usepackage{subfigure}
\usepackage{natbib}
\usepackage{epsfig}
\usepackage{amsfonts}
\usepackage{mathrsfs}
\usepackage{CJK}
\usepackage[toc,page,title,titletoc,header]{appendix}

\begin{document}
\title{Dynamical speedup of a two-level system induced by coupling in the hierarchical environment}

\author{Kai Xu}
 \affiliation{Key Laboratory of Micro-Nano Measurement-Manipulation and Physics (Ministry of
Education), School of Physics and Nuclear Energy Engineering, Beihang University,
Xueyuan Road No. 37, Beijing 100191, China}

\author{Guo-Feng Zhang}
\email{gf1978zhang@buaa.edu.cn}
 \affiliation{Key Laboratory of Micro-Nano Measurement-Manipulation and Physics (Ministry of
Education), School of Physics and Nuclear Energy Engineering, Beihang University,
Xueyuan Road No. 37, Beijing 100191, China}

\author{Wu-Ming Liu}
 \affiliation{Beijing National Laboratory for Condensed Matter Physics, Institute of Physics, Chinese Academy of Sciences, Beijing, China}

\date{\today}
\begin{abstract}
We investigate the dynamics of a two-level system in the presence of an
overall environment composed of two layers. The first layer is just one
single-mode cavity which decays to memoryless reservoir while the second
layer is the two coupled single-mode cavities which decay to memoryless
or memory-keeping reservoirs. In the weak-coupling regime between the qubit
and the first-layer environment, our attention is focused on the effects of
the coupling in the hierarchical environment on the non-Markovian speedup
dynamics behavior of the system. We show that, by controlling the coupling
in the second-layer environment, the multiple dynamics crossovers from
Markovian to non-Markovian and from no-speedup to speedup can be realized.
This results hold independently on the nature of the second-layer environment.
Differently, we find that how the coupling between the two layers affects the
non-Markovian speedup dynamics behavior depends on the nature of the second-layer
environment.
\end{abstract}

\maketitle

\section{\textbf{{Introduction}}}

Recently, motivated by real physical systems, understanding the evolution
of open quantum systems that are coupled to the environment has drawn
more and more interest \cite{Caruso1,Lo Franco2,Rivas3,Lee4,Cederbaum5}.
In general, such physical systems can interchange with its environment
particles, energy or information and therefore lead to decoherence and
dissipation of the system's quantum properties \cite{Petruccione}.
In this situation, a monotonic one-way flow of information from the system to
the environment tends to the appearance what is called Markovian dynamics.
However, in many scenarios, due to the increasing capability to manipulate
quantum systems, leading to the occurrence of a backflow of information from the
environment to the system, what is called non-Markovian dynamics \cite{Vega6,
Vacchini7,Lorenzo,Plenio,Caruso,Laine,Dajka,Smirne,Luo,Luxm,Chruscinski2,SWibmann,
Lijg,Alimm,Wang0,Smirne0,Franco00,Laine2,Liu4,Man1,Man2}. The non-Markovian effects
not only suppress the decay of the coherence or the entanglement of quantum systems \cite{Zou,Addis}
but also play a leading role in many real physical processes such as quantum state engineering,
quantum control \cite{Xue,D'Arrigo9,Bylicka} and the quantum information processing
\cite{Xiang,Bennett,Xu11,Aaronson,Duan}. For example, recent experiments \cite{sunfangwen}
have shown that non-Markovianity can improve the probability of success of the
Deutsch-Jozsa algorithm in diamonds.

The non-Markovianity has received great attention in the form of quantitative measurement
\cite{Piilo0,Plastina,Huelga0,Laine0,Chru?ci¨½ski0,Maniscalco0,Hall0}, experimental demonstration
\cite{Chiuri0,Huang,Bernardes} and the impact on the speedup evolution of quantum system
\cite{Fan,Sun,Meng,Zhangyj3,Xuzy,Anjh2,Zhangyj4,Xuzy2,Cianciaruso,Deffner}. For instance,
a new characterization of non-Markovian quantum evolution based on the concept of non-Markovianity
degree has been proposed \cite{Piilo0}. The experimental realisation of a non-Markovian process
where system and environment are coupled through a simulated transverse Ising model has been reported \cite{Chiuri0}.
The non-Markovian effect could induce speedup dynamics process in the strong system-environment
coupling regime for the damped Jaynes-Cummings model \cite{Deffner}. And this novel has been realized by
increasing the system-environment coupling strength and the number of atoms in a controllable
environment \cite{Cimmarusti}.

In the previous studies, some researchers have considered the quantum system coupled to a
single-layer environment. However, the system can be influenced by multilayer environments
\cite{Hanson1,Hanson2,Pla,Chekhovich,Tyryshkin} in the realistic scenarios. For example,
the electron spin in a quantum dot may be influenced strongly by the
surrounding nuclei \cite{Hanson1}. The surrounding nitrogen impurities constitute the principal
bath for a nitrogen-vacancy center, while the carbon-13 nuclear spins also have some
influences on it \cite{Hanson2,Chekhovich}. Based on these, multilayered environments have
been considered for the study of non-Markovian dynamics of the system. A qubit
that is coupled to a hierarchical environment, which contains a single-mode cavity and a
reservoir consisting of an infinite number of modes has been investigated \cite{Mat}.  They
show the non-Markovian character of the system is influenced by the coupling strength
between the qubit and cavity and the correlation time of the reservoir. Besides,
the hierarchical environment model where the first layer is just a single lossy cavity
while the second layer consists of a number of lossy cavities has been considered
\cite{Nguyen}. In this model, the increase the number of the lossy cavities and the coupling between
the two layers can trigger the non-Markovian dynamics of the system.
However, in the experiment, the coupling relationship between different parts of a complex environment
has a great influence on the dynamic behavior of the system. For example, a single spin interacting with
an adjustable spin bath shows that both the internal interactions of the bath and the coupling between
the central spin and the bath can be tuned in situ, allowing access to regimes with surprisingly different
behavior \cite{Hanson1}. So in the treatments of the composite environments, the influence of the coupling
relationship between various parts of the overall environment on the dynamic behavior of the system has
to be taken into account.

Based on these, we mainly investigate the dynamics of a qubit coupled with the overall environment composed of two layers.
By using the quantum speed limit (QSL) time \cite{Deffner,8,08}
to define the speedup evolutional process, the influence of the coupling in
the second-layer environment and the coupling between the two layers on the quantum
evolutional speed of the qubit are discussed in the weak-coupling regime between
the qubit and the first-layer environment. In this paper, by considering the second-layer environment with different properties, we elaborate how the
non-Markovian speedup dynamics of the system can be obtained by controlling the
coupling in the the hierarchical environment. Besides,
it also explains the process that the coupling in the hierarchical environment
affects the non-Markovian speedup dynamics of the system when the second-layer environment has different properties.

\section{\textbf{Theoretical model}}

\begin{figure}[tbh]
\includegraphics*[bb=124 115 573 506,width=8cm, clip]{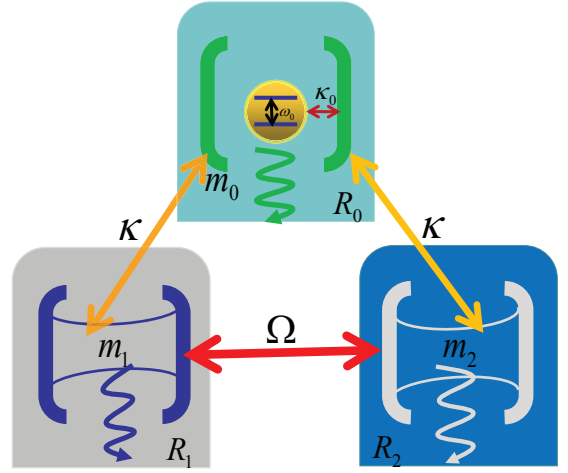}
\caption{Schematic representation of the model.
A two-level system is coupled with strength $\kappa_0$ to a mode
$m_0$ which decays to a memoryless reservoir $R_{0}$ with a lossy rate $\Gamma_{0}$.
The mode $m_0$ is further coupled simultaneously with strengths
$\kappa$ to modes $m_{1}$, $m_{2}$ which also decay to their respective
reservoirs $R_{1}$, $R_{2}$ with lossy rates $\Gamma_{1,2}=\Gamma$.
The two cavity modes $m_{1}$ and $m_{2}$ are coupled with
strength $\Omega$.}
\end{figure}

We consider that the entire system consists of a qubit and
the hierarchical environment where the cavity $m_{0}$ and its
corresponding memoryless reservoir $R_{0}$ serve as the first-layer
environment for the two-level system and the two coupled
cavities $m_{1}$, $m_{2}$ and reservoirs $R_{1}$, $R_{2}$ involved act as the
second-layer environment, as depicted in Fig. 1. More precisely,
the two-level system is coupled with strength $\kappa_0$ to
the mode $m_0$ which decays to a memoryless reservoir $R_{0}$
with a lossy rate $\Gamma_{0}$ and then the mode $m_0$ is
further coupled with strength $\kappa$ to the modes $m_1$, $m_2$
which decay to their respective reservoirs
$R_1$, $R_2$ with lossy rates $\Gamma_{1,2}=\Gamma$. Furthermore, the coupling
strength between cavities $m_1$ and $m_2$ is $\Omega$.
For the sake of simplicity, we assume that the frequency $\omega_{n}$
of the mode $m_n$ is equal to the qubit transition
frequency $\omega_{0}$, i.e., $\omega_{n}=\omega_{0}$.
The total Hamiltonian is given by $H=H_{0}+H_{I}$, reads

\begin{eqnarray}
H_{0}&=&\frac{\omega_{0}}{2}\sigma_{z}+\sum_{n=0}^{2}\omega_{n}b^{\dag}_{n}b_{n}+\sum_{n=0}^{2}\sum_{k}\omega_{n,k}c^{\dag}_{n,k}c_{n,k},\nonumber\\
H_{I}&=&\kappa_{0}(\sigma_{+}b_{0}+\sigma_{-}b^{\dag}_{0})+\sum_{n=1}^{2}\kappa(b_{0}b^{\dag}_{n}+b^{\dag}_{0}b_{n})\nonumber\\
&+&\Omega(b^{\dag}_{1}b_{2}+b^{\dag}_{2}b_{1})+\sum_{n=0}^{2}\sum_{k}g_{n,k}(b_{n}c^{\dag}_{n,k}+b^{\dag}_{n}c_{n,k}). \label{01}
\end{eqnarray}
In Eq (1), $\omega_{0}$ is the transition frequency of the qubit system,
$\sigma_{\pm}$ denote the raising and lowering operators of the
qubit, $b^{\dag}_{n}$ ($b_{n}$) is the
bosonic creation (annihilation) operators for the mode $m_{n}$ with
frequency $\omega_{n}$, while $\kappa_{0}$, $\kappa$, $\Omega$
represent the corresponding couplings. Furthermore, $c^{\dag}_{n,k}$
($c_{n,k}$) is the creation (annihilation) operator of field mode
$k$ with frequency $\omega_{n,k}$ of reservoir $R_{n}$ and $g_{n,k}$ denotes the
coupling of the mode $m_{n}$ with the the mode $k$ of its own reservoir $R_{n}$.
In the interaction picture, Eq. (1) can be written as

\begin{eqnarray}
H_{int}&=&\kappa_{0}(\sigma_{+}b_{0}+\sigma_{-}b^{\dag}_{0})+\sum_{n=1}^{2}\kappa(b_{0}b^{\dag}_{n}+b^{\dag}_{0}b_{n})\nonumber\\
&+&\Omega(b^{\dag}_{1}b_{2}+b^{\dag}_{2}b_{1})+\sum_{n=0}^{2}\sum_{k}g_{n,k}(b_{n}c^{\dag}_{n,k}e^{i\Delta_{n,k}t}\nonumber\\
&+&b^{\dag}_{n}c_{n,k}e^{-i\Delta_{n,k}t}),\label{02}
\end{eqnarray}
where $\Delta_{n,k}=\omega_{n,k}-\omega_{0}$.

The reservoirs $R_{1}$ and $R_{2}$ in the second layer environment may be
memoryless or memory-keeping. Selecting different
types of the reservoir correspond to different methods to obtain the dynamic
evolution of the qubit. Then as we all known, if there is no second layer
environment of our system, the dynamics of the qubit depends on the parameters
$k_{0}$ and $\Gamma_{0}$ in such a way that $k_{0}<\Gamma_{0}/4$ ($k_{0}>\Gamma_{0}/4$),
identified as the weak (strong) coupling regime. Below, in the weak qubit-$m_{0}$ coupling regime,
we take the second-layer environment with different properties as examples to study the dynamical evolution of the system.

\section{\textbf{memoryless nature of the second-layer environment}}

In the section, we consider the reservoirs $R_1$ and $R_2$ in
the second layer environment are Markovian (memoryless). In other words,
the correlation times of the reservoirs $R_{1}$ and $R_{2}$ are
much smaller than the single-mode relaxtion time. In this case,
the density operator $\rho$ of the total system is

\begin{eqnarray}
\frac{d\rho}{dt}=&-&i[H,\rho]-\frac{\Gamma_{0}}{2}(b_{0}^{\dag}b_{0}\rho-2b_{0}{\rho}b^{\dag}_{0}+{\rho}b^{\dag}_{0}b_{0})\nonumber\\
&-&\sum_{n=1}^{2}\frac{\Gamma_{n}}{2}(b^{\dag}_{n}b_{n}\rho-2b_{n}{\rho}b^{\dag}_{n}+{\rho}b^{\dag}_{n}b_{n}) \label{03},
\end{eqnarray}
where $\Gamma_{0}$ and $\Gamma_{n}$ denote the dissipation
rate of the cavities $m_{0}$ and $m_{n}$,
respectively. For simiplicity, we assume that the qubit
is initially in the excited state $|1{\rangle}_{s}$ and
three modes are in the ground state $|0{\rangle}_{s}$,i.e.,
the total system initial state is $\rho(0)=|1000{\rangle}{\langle}1000|$
with $\psi(0)=|1000{\rangle}$. Since there exists at most
one excitation in the total system at any time, then at time $t$ the
evolutional state of the total system can be written as
$|\psi(t){\rangle}=a(t)|1000{\rangle}+c_{0}(t)|0100{\rangle}+c_{1}(t)|0010{\rangle}
+c_{2}(t)|0001{\rangle},$ where $a(t)$, $c_{0}(t)$, $c_{1}(t)$, $c_{2}(t)$ correspond to
probability amplitudes of the excited state for the atom or the
modes $m_{0}$, $m_{1}$, $m_{2}$ with $a(0)=1$ and $c_{0}(0)=c_{1}(0)=c_{2}(0)=0$.
Besides, the probability amplitudes $a(t)$, $c_{0}(t)$, $c_{1}(t)$,$c_{2}(t)$
are governed by the Hamiltonians in Eq. (\ref{01}), and determined by a set of
differential equations as

\begin{eqnarray}
\begin{aligned}
i\dot{a}(t)&=\kappa_{0}c_{0}(t), \\
i\dot{c_{0}}(t)&=-\frac{i}{2}\Gamma_{0} c_{0}(t)+\kappa_{0}a(t)+{\kappa}(c_{1}(t)+c_{2}(t)),\\
i\dot{c_{1}}(t)&=-\frac{i}{2}\Gamma c_{1}(t)+\kappa c_{0}(t)+\Omega c_{2}(t),\\
i\dot{c_{2}}(t)&=-\frac{i}{2}\Gamma c_{2}(t)+\kappa c_{0}(t)+\Omega c_{1}(t).
\label{04}
\end{aligned}
\end{eqnarray}

The solutions of the above equations can be obtained by Laplace
transformation and Laplace inverse transformation combined with
numerical simulation. Then the reduced density matrix of the qubit in the
atomic basis $\{|1{\rangle}_{s}$, $|0{\rangle}_{s}\}$ can be
expressed as
\begin{equation}
\rho^{s}(t)=\left(
  \begin{array}{cc}
    \rho_{11}(0)|a(t)|^{2} & \rho_{01}(0)a(t)^{\ast} \\
    \rho_{10}(0) a(t)      & \rho_{00}(0)+\rho_{11}(0)(1-|a(t)|^{2})\label{05} \\
  \end{array}
\right)
\end{equation}
where $\rho_{11}(0)=1$, $\rho_{00}(0)=\rho_{01}(0)=\rho_{10}(0)=0$.

\begin{figure}[tbh]
\includegraphics*[bb=36 19 282 361,width=8cm, clip]{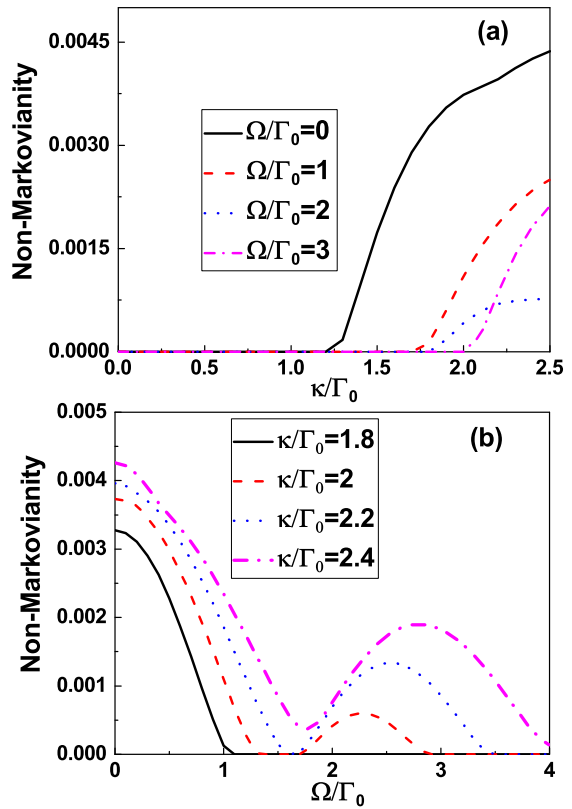}
\caption{(Color online) (a), (b) The non-Markovianity
$\mathbf{N}(\Phi)$ of the atomic system dynamics process from
$\rho^{s}_{0}=|1\rangle_{s}{\langle}1|$ to $\rho^{s}_{\tau}$ as a
function of the coupling strength $\kappa$ between the two layers
and the coupling strength $\Omega$ in the second-layer environment with memoryless effects
for the weak qubit-$m_{0}$ coupling regime (i.e, $\kappa_{0}=0.2\Gamma_{0}$). The
parameters are: $\Gamma=\Gamma_{0}$, $\tau=4$.}
\end{figure}

A measure $\mathbf{N}(\Phi)$ of non-Markovianity based on
the distinguishability between the evolutions of two different
initial states of the system has been defined by Breuer $et$ $al$. \cite{Piilo0}.
For a quantum process $\Phi(t)$, $\rho^{s}(t)=\Phi(t)\rho^{s}(0)$, with $\rho^{s}(0)$ and
$\rho^{s}(t)$ denote the density operators at time $t=0$ and at any
time $t>0$ of the quantum system, respectively, this suggests defining the measure
$\mathbf{N}(\Phi)$ for the non-Markovianity of the quantum process $\Phi(t)$ through
$\mathbf{N}(\Phi)=\max_{\rho^{s}_{1,2}(0)}\int_{\sigma>0}dt\sigma[t,\rho^{s}_{1,2}(0)],
$ with
$\sigma[t,\rho^{s}_{1,2}(0)]=\frac{d}{dt}\mathcal{D}(\rho^{s}_{1}(t),\rho^{s}_{2}(t))$
is the rate of change of the trace distance.
$\mathcal{D}(\rho^{s}_{1},\rho^{s}_{2})=\frac{1}{2}\|\rho^{s}_{1}-\rho^{s}_{2}\|,
$ where $\|M\|=Tr(\sqrt{M^{\dag}M})$ and $0{\leq}\mathcal{D}\leq1$.
And $\sigma[t,\forall\rho^{s}_{1,2}(0)]\leq0$
corresponds to all dynamical semigroups and all time-dependent
Markovian processes. To evaluate the non-Markovianity, we should find
a specific pair of optimal initial states $\rho^{s}_{1,2}(0)$ to maximize
the time derivative of the trace distance. The states of these optimal pairs must
be orthogonal and lie on the boundary of the space of physical states\cite{SWibmann}.
Through a numerical simulation, it is proven that the optimal
state pair of the initial states can be chosen as
$\rho^{s}_{1}(0)=(|0\rangle_{s}+|1\rangle_{s})/\sqrt{2}$ and
$\rho^{s}_{2}(0)=(|0\rangle_{s}-|1\rangle_{s})/\sqrt{2}$
\cite{Lijg,Piilo0}. Here, for our model, by selecting this optimal state pair,
the rate of change of the trace distance can be derived in a simple form as
$\sigma[t,\rho^{s}_{1,2}(0)]=\partial_{t}|a(t)|$.
Then the non-Markovianity of the quantum system dynamics process
from $\rho^{s}(0)$ to $\rho^{s}(t)$ can be calculated by
$\mathbf{N}(\Phi)=\int^{t}_{0}[\partial_{t}|a(t)|]|_{>0}dt$.

\begin{figure}[!tbh]
\includegraphics*[bb=1 2 395 347,width=8cm, clip]{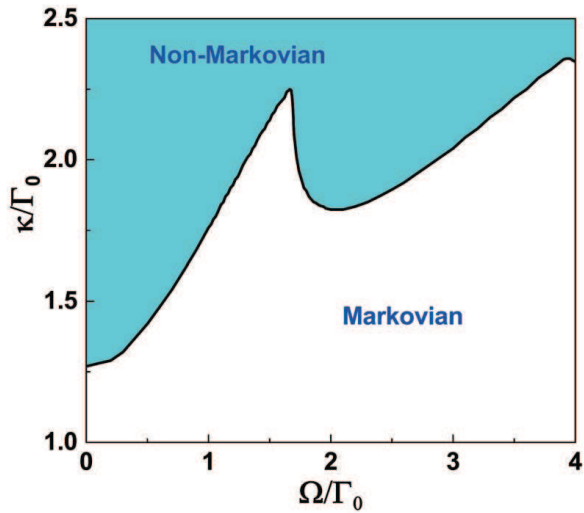}
\caption{(Color online) Phase diagram of the non-Markovianity
$\mathbf{N}(\Phi)$ of the atomic system dynamics process from
$\rho^{s}_{0}=|1\rangle_{s}{\langle}1|$ to $\rho^{s}_{\tau}$
in the $\kappa/\Gamma_{0}$-$\Omega/\Gamma_{0}$ plane
with $\kappa_{0}=0.2\Gamma_{0}$ in the weak qubit-$m_{0}$ coupling
regime. The parameter is $\tau=4$.}
\end{figure}

Without the other-layer environment, the system experiences the Markovian dynamics in the weak coupling regime
($\Gamma_{0}>4\kappa_{0}$).
In the case of adding the second-layer environment with different properties,
the atomic dynamics process from $\rho^{s}_{0}$ to
$\rho^{s}_{\tau}$ would be considered in the weak qubit-$m_{0}$
coupling regime (here $\tau$ is the actual evolution time).
Firstly, in the case of the second-layer environment with the memoryless nature,
the non-Markovianity of the atom dynamics as function of the controllable
hierarchical environment parameters ($\kappa$, $\Omega$) has been plotted in Fig. 2.
By fixing $\Omega/\Gamma_{0}$ in Fig. 2(a),
a remarkable dynamical crossover from Markovian behavior to non-Markovian behavior can
occur at a certain critical coupling strength $\kappa_{c}$.
When $\kappa<\kappa_{c}$, the system exhibits Markovian dynamics behavior,
and then the non-Markovianity increases monotonically with increasing $\kappa/\Gamma_{0}$.
Differently, the variations of the non-Markovianity can be abundant
with respect to the scaled coupling strength $\Omega/\Gamma_{0}$, as shown in Fig. 2(b).
For relatively small values of $\kappa$ (e.g., $\kappa=1.8\Gamma_{0}$), the non-Markovianity
decreases monotonically with increasing $\Omega$, and then the dynamics abides Markovian behavior.
For larger values of $\kappa$, the non-Markovianity of the atom dynamics can be nonmonotonic as
increasing $\Omega$. In this case, the non-Markovianity can experience successive decreasing and
increasing behaviors with increasing $\Omega$. For particular values of
the coupling strength between the two layers (e.g., $\kappa=2\Gamma_{0}$, $2.2\Gamma_{0}$),
the non-Markovianity may vanish within a finite
interval of $\Omega$ and revive again, eventually vanishing at relatively large $\Omega$.
In other words, the successive transitions between non-Markovian and Markovian regimes for
the system dynamics can be induced by controlling the coupling strength $\Omega$ in the second-layer environment.
Furthermore, by fixing $\kappa/\Gamma_{0}$
in Fig. 2(b), we can be surprised to find that, compared with the absence of coupling in the second-layer environment
(i.e., $\Omega=0$), the introduction of coupling between the modes $m_{1}$ and $m_{2}$ cannot
enhance the non-Markovianity of the system.
Finally, it needs to be emphasized that, by considering the second-layer environment
with the memoryless nature in the weak qubit-$m_{0}$ coupling regime,
manipulations the coupling strength in the hierarchical environment can trigger the non-Markovian dynamics behavior of the system.

To comprehensively understand the impacts of the hierarchical environment parameters mentioned above in the weak qubit-$m_{0}$
coupling regime, Fig. 3 describes the $\kappa$-$\Omega$ phase diagrams of the
transitions between the non-Markovian and Markovian dynamics. It is clear that, by fixing $\Omega$,
the crossover between Markovian and non-Markovian dynamics can be occurred as increasing $\kappa$.

\section{\textbf{memory-keeping nature of the second-layer environment}}

In the previous section, we have considered the qubit is coupled to
a mode $m_{0}$ which decays to a memoryless reservoir
and then the mode $m_{0}$ is interacting with two coupled modes, $m_{1}$ and $m_{2}$,
which are dissipated respectively by two memoryless reservoirs $R_{1}$ and
$R_{2}$. We have already known that in the weak qubit-$m_{0}$ coupling
regime, three modes in the hierarchical environment assume full
responsibility for arousing the environmental memory effect. However,
it is necessary to study the overall memory effects of the two-level system
due to the coupling changes if these three modes are only part of the
overall environmental memory. Based on this, in the weak qubit-$m_{0}$ coupling regime,
we consider the modes $m_{1}$ and $m_{2}$ in the second-layer environment
are dissipated by structured reservoirs $R_{1}$, $R_{2}$ exhibiting memory effects.

In the following, we assume the qubit is initially in the excited state while the three
modes and the corresponding to reservoirs are in their ground state. So the evolutional
initially state of the total system is $|\varphi(0)\rangle=|1\rangle_{s}|000\rangle_{m_{0}m_{2}m_{3}}|\overline{000}\rangle_{R_{0}R_{1}R_{2}}$ with
$|\overline{0}\rangle_{R_{n}}=\prod_{k}|\overline{0_{k}}\rangle_{R_{n}}$. Then the evolution of the total system at time $t$ can be given as

\begin{eqnarray}
|\varphi(t){\rangle}&=&h(t)|1{\rangle}_{s}|000{\rangle}_{m_{0}m_{1}m_{2}}|\overline{000}{\rangle}_{R_{0}R_{1}R_{2}}\nonumber\\
&+&|0{\rangle}_{s}(c_{0}(t)|100{\rangle}+c_{1}(t)|010{\rangle}\nonumber\\
&+&c_{2}(t)|001{\rangle})_{m_{0}m_{1}m_{2}}|\overline{000}{\rangle}_{R_{0}R_{1}R_{2}}\nonumber\\
&+&|0{\rangle}_{s}|000{\rangle}_{m_{0}m_{1}m_{2}}\sum_{n=0}^{2}\sum_{k}c_{n,k}|1_{k}{\rangle}_{R_{n}}|\overline{0}{\rangle}_{R_{\overline{n}}}\label{04},
\end{eqnarray}
where $|1_{k}{\rangle}_{R_{n}}$=$|00\cdots1_{k}\cdots00{\rangle}_{R_{n}}$ denotes that one excitation in the $kth$ mode of the reservoir $R_{n}$ and $\overline{n}$ is complementary to $n$.  And when $t=0$, $h(0)=1,c_{0}(0)=c_{1}(0)=c_{2}(0)=c_{n,k}(0)=0$. Besides, the non-Hermitian Hamiltonian, which includes the additional terms $-\frac{i\Gamma_{0}a^{+}a}{2}$, is considered to approximate the dissipative effect of the lossy cavity $m_{0}$. Put Eq. (6) into the Schr$\ddot{o}$dinger equation, the Hamiltonian of the total system in the interaction picture is determined by the following equations

\begin{eqnarray}
\begin{aligned}
\dot{h}(t)&=-i\kappa_{0}c_{0}(t), \\
\dot{c_{0}}(t)&=-\frac{1}{2}\Gamma_{0} c_{0}(t)-i\kappa_{0}h(t)-i{\kappa}(c_{1}(t)+c_{2}(t)),\\
\dot{c_{1}}(t)&=-i\kappa c_{0}(t)-i\Omega c_{2}(t)-ig_{1,k}e^{i\Delta_{1,k}t} c_{1,k}(t),\\
\dot{c_{2}}(t)&=-i\kappa c_{0}(t)-i\Omega c_{1}(t)-ig_{2,k}e^{i\Delta_{2,k}t} c_{2,k}(t),\\
\dot{c_{0,k}}(t)&=0,\\
\dot{c_{1,k}}(t)&=-ig_{1,k}e^{i\Delta_{1,k}t} c_{1}(t),\\
\dot{c_{2,k}}(t)&=-ig_{2,k}e^{i\Delta_{2,k}t} c_{2}(t).
\label{04}
\end{aligned}
\end{eqnarray}
Integrate the last two equations above with the initial condition $c_{n,k}=0$ and
bring the results into the third and fourth equations above, we get the following equation

\begin{eqnarray}
\begin{aligned}
\dot{c_{1}}(t)&=-i\kappa c_{0}(t)-i\Omega c_{2}(t)\\
&-\int_{0}^{t}\Sigma_{k}|g_{1,k}|^{2}e^{-i\Delta_{1,k}(t-t^{'})}c_{1}(t^{'})dt^{'},\\
\dot{c_{2}}(t)&=-i\kappa c_{0}(t)-i\Omega c_{1}(t)\\
&-\int_{0}^{t}\Sigma_{k}|g_{2,k}|^{2}e^{-i\Delta_{2,k}(t-t^{'})}c_{2}(t^{'})dt^{'}.
\label{04}
\end{aligned}
\end{eqnarray}
The $\Sigma_{k}|g_{n,k}|^{2}e^{-i\Delta_{n,k}(t-t^{'})}$ in the above equation is
recognized as a correlation function $f_{n}(t-t^{'})$ of the reservoir $R_{n}$ in the second-layer environment.
The correlation function $f_{n}(t-t^{'})=\int d\omega J_{n}(\omega) e^{i(\omega_{0}-\omega)(t-t^{'})}$
is related to the spectral density $J_{n}(\omega)$ of the reservoir $R_{n}$. Now we
suppose that the reservoir $R_{n}$ ($n=1,2$) has a Lorentzian spectral
$J_{n}(\omega)=\frac{\Upsilon_{n}\lambda_{n}^{2}}{2\pi((\omega-\omega_{0})^{2}+\lambda_{n}^{2})}$,
where the coupling strength between the mode $m_{n}$ and the reservoir $R_{n}$ is $\Upsilon_{n}$ and the
correlation time of the reservoir $R_{n}$ is $\lambda_{n}^{-1}$. Then the two-point correlation function of the
reservoir $R_{n}$ in the second-layer environment can be written
as $f_{n}(\tau)=\frac{1}{2}\Upsilon_{n}\lambda_{n}e^{-\lambda_{n}|\tau|}$. The solutions of the amplitude
$h(t)$ can be obtained by solving Eqs. (7) and (8). Then we can analysis the dynamical evolution of the system.

\begin{figure}[tbh]
\includegraphics*[bb=1 0 245 399,width=8cm, clip]{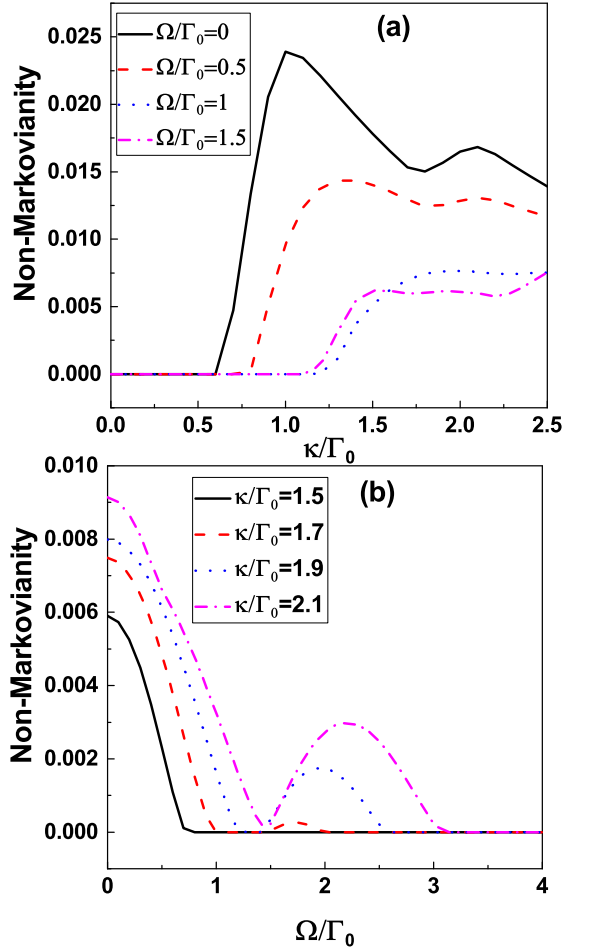}
\caption{(Color online) (a), (b) The non-Markovianity
$\mathbf{N}(\Phi)$ of the atomic system dynamics process from
$\rho^{s}_{0}=|1\rangle_{s}{\langle}1|$ to $\rho^{s}_{\tau}$ as a
function of the coupling strength $\kappa$ between the two layers
and the coupling strength $\Omega$ in the second-layer environment with memory
effects for the weak qubit-$m_{0}$ coupling regime (i.e., $\kappa_{0}=0.2\Gamma_{0}$). The
other parameters are: (a) $\Upsilon_{1}=\Upsilon_{2}=\Gamma_{0}$, $\lambda_{1}=\lambda_{2}=0.1\Gamma_{0}$, $\tau=4$; (b) $\Upsilon_{1}=\Upsilon_{2}=\Gamma_{0}$, $\lambda_{1}=\lambda_{2}=\Gamma_{0}$, $\tau=4$.}
\end{figure}

In the weak qubit-$m_{0}$ coupling regime, Figs. 4(a) and 4(b) show
how the non-Markovianity of the atomic dynamics
is affected by the hierarchical environment parameters $\Omega$
or $\kappa$ when the second-layer
environment has a memory-keeping effect. It is worth noting that, by fixing $\Upsilon_{1}=\Upsilon_{2}=\Gamma_{0}$,
$\lambda_{1}=\lambda_{2}=0.1\Gamma_{0}$ in the Fig. 4(a), the dynamical crossover
from Markovian behavior to non-Markovian behavior for the atomic dynamics ($\rho^{s}_{0}=|1\rangle_{s}{\langle}1|$
to $\rho^{s}_{\tau}$) could appear as increasing coupling strength $\kappa$.
And for a given $\Omega$, we find that the
non-Markovianity can be nonmonotonic with respect to the coupling strength $\kappa$.
This implies the increases of the coupling strength $\kappa$ can not only enhance the non-Markovianity of
the system, but also weaken it. Besides,
when $\kappa=1.5\Gamma_{0}$ is relative
small in the Fig. 4(b), the non-Markovianity monotonically decreases with increasing $\Omega$, and then the system always remains
Markovian dynamics behavior.
While $\kappa$ is larger (e.g., $\kappa=1.7\Gamma_{0}$, $1.9\Gamma_{0}$, $2.1\Gamma_{0}$), the increase
of $\Omega$ can induce continuous
increase and decrease behaviors of the non-Markovianity.
Finally, it is worth noting that, compared with the memoryless effect in the second-layer environment,
the introduction of memory-keeping effect gives $\kappa$ a double impact (i.e., the enhancement of $\kappa$
can improve and weaken non-Markovianity).

\section{\textbf{{Quantum speedup of the atomic dynamics}}}

In this section, we will use the definition of the QSL time for
an open quantum system, which can be helpful to analyze the maximal
speed of evolution of an open system. The QSL time between an initial state
$\rho^{s}(0)=|\phi_{0}\rangle\langle\phi_{0}|$ and its target state
$\rho^{s}(\tau)$ (the evolutional state of the system at the
actual evolution time $\tau$) for open system is defined by
\cite{Deffner}
$\tau_{QSL}=\sin^{2}[\mathbf{B}(\rho^{s}(0),\rho^{s}(\tau))]/\Lambda^{\infty}_{\tau}$,
where
$\mathbf{B}(\rho^{s}(0),\rho^{s}(\tau))=\arccos\sqrt{\langle\phi_{0}|\rho^{s}(\tau)|\phi_{0}\rangle}$
denotes the Bures angle between the initial and target states of the
system, and
$\Lambda^{\infty}_{\tau}=\tau^{-1}\int^{\tau}_{0}\|\dot{\rho}^{s}(t)\|_{\infty}dt$
with the operator norm $\|\dot{\rho}^{s}(t)\|_{\infty}$ equaling to
the largest singular value of $\dot{\rho}^{s}(t)$.
$\tau_{QSL}/\tau=1$ means the quantum system evolution is already along
the fastest path and possesses no potential capacity for further
quantum speedup. While for the case $\tau_{QSL}/\tau<1$, the speedup evolution of
the quantum system may occur and the much shorter $\tau_{QSL}/\tau$, the
greater the capacity for potential speedup will be.

In the light of Eq. (\ref{05}), the relationship between non-Markovianity and the QSL
times has been given by  $\frac{\tau_{QSL}}{\tau}
=\frac{1-|a(\tau)|^{2}}{2\mathbf{N}(\Phi)+1-|a(\tau)|^{2}}$
\cite{Xuzy,Zhangyj4}. The
above equation means that the larger the non-Markovianity would
lead to the lower the QSL times (that is to say, the greater the
capacity for potential speedup could be). In this model, by the controllable non-Markovianity discussed above,
the speedup of the quantum system can also be achieved.
Below, in the case of the second-layer environment with the different properties,
we mainly focus on the influence of the coupling in the hierarchical environment on the dynamical speedup
of quantum system in the weak qubit-$m_{0}$ coupling regime.

\begin{figure}[tbh]
\includegraphics*[bb=0 49 513 528,width=8cm, clip]{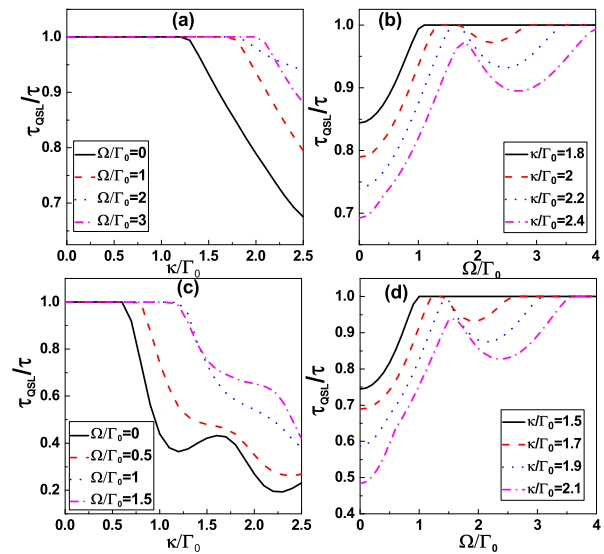}
\caption{(Color online) (a), (b) The QSL time for the atomic system
dynamic process as a function of the coupling strength
$\kappa$ between the two layers and the coupling strength $\Omega$ in the
second-layer environment with memoryless effects.
(c), (d) The QSL time for the atomic system
dynamic process as a function of the coupling strength
$\kappa$ between the two layers and the coupling strength $\Omega$ in the
second-layer environment with memory effects.
The other parameters are : (a), (b) $\kappa_0=0.2\Gamma_{0}$, $\Gamma=\Gamma_{0}$;
(c) $\kappa_0=0.2\Gamma_{0}$, $\lambda_{1}=\lambda_{2}=0.1\Gamma_{0}$, $\Upsilon_{1}=\Upsilon_{2}=\Gamma_{0}$;
(d) $\kappa_0=0.2\Gamma_{0}$, $\lambda_{1}=\lambda_{2}=\Gamma_{0}$, $\Upsilon_{1}=\Upsilon_{2}=\Gamma_{0}$, $\kappa_0=0.2\Gamma_{0}$.}
\end{figure}

For the second-layer environment with memoryless nature in the weak qubit-$m_{0}$ coupling regime,
the QSL time $\tau_{QSL}/\tau$ as a function of the coupling strength $\Omega$ and $\kappa$ have
been plotted in Figs. 5 (a) and (b). By fixing $\Omega/\Gamma_{0}$ in the Fig. 5(a),
the no-speedup evolution ($\tau_{QSL}/\tau=1$) could be followed, and
the speedup evolution ($\tau_{QSL}/\tau<1$) would occur when the
coupling strength $\kappa$ is larger than a certain critical coupling
strength $\kappa_{c1}$. As for Fig. 5(b), by fixing $\kappa=1.8\Gamma_{0}$, the speedup evolution of the system can be induced
by decreasing $\Omega$. However, in the case $\kappa=2.4\Gamma_{0}$, no matter
how we adjust the parameter $\Omega$, the system always remains speedup evolution. Besides, when the coupling strength $\kappa$
takes the particular values (e.g., $\kappa=2\Gamma_{0}$, $2.2\Gamma_{0}$), the quantum system can experience successive transforms
from speedup evolution to no-speedup evolution as increasing $\Omega$. This means the speed of evolution
for the system can be controlled to a speed-up or speed-down process by manpulating the coupling strength $\Omega$.
So in the case of the second-layer environment with the memoryless nature in the weak qubit-$m_{0}$
coupling regime, the purpose of accelerating evolution can be achieved by controlling the coupling strength in the hierarchical environment.

For the second-layer environment with memory-keeping nature in the weak qubit-$m_{0}$ coupling regime,
the variation of the QSL time with respect to $\kappa$ and $\Omega$ have been shown in Figs. 5 (c) and (d).
When the value $\Omega$ is confirmed in Fig. 5(c), in the case $\kappa<\kappa_{c2}$ ($\kappa_{c2}$ means the
critical value of $\kappa$), the QSL time is always equal to the actual evolution time, but the QSL time can
be nonmonotonic with increasing $\kappa$ when $\kappa>\kappa_{c2}$. That is to say, by manipulating $\kappa$,
the evolution of the system can be accelerated and its evolution speed can also be controlled.
Besides, by confirming $\kappa/\Gamma_{0}=1.5$ in Fig. 5(d), the QSL time monotonically increases as increasing $\Omega$ and then
remains at the actual evolution time. This means that the speedup evolution may appear as decreasing $\Omega$.
However, in the case $\kappa=1.7\Gamma_{0}$, $1.9\Gamma_{0}$, $2.1\Gamma_{0}$, the QSL time nonmonotonically increases
with increasing $\Omega$ and eventually remains in the actual evolution time. These behaviors are similar to the
second-layer environment with the memoryless nature, one obvious difference is that the introduction of memory effects
makes it possible to control the speed (speed-up or speed-down) of evolution for the system by manipulating $\kappa$.

\section{\textbf{{Conclusion}}}

In conclusion, for the weak qubit-$m_{0}$ coupling regime, we have
investigated the dynamics behavior of the qubit in a controllable hierarchical
environment where the first-layer environment is a mode $m_{0}$ which
decays to a memoryless reservoir and the second-layer environment is
two coupled modes $m_{1}$ and $m_{2}$ which decay to memoryless
or memory-keeping reservoirs.
In the case of the memoryless nature of the second-layer environment,
the three modes ($m_{0},m_{1},m_{2}$) can be regarded as the memory source
of the overall environment. In the case of the memory-keeping nature
of the second-layer environment, the three modes can only be a part of the total
memory source of the overall environment. In the above two cases, by controlling the
coupling strength $\kappa$ between the two layers and the coupling strength $\Omega$ in
the second-layer environment, two dynamical crossovers
of the quantum system, from Markovian to non-Markovian dynamics and from no-speedup
evolution to speedup evolution, have been achieved in the weak qubit-$m_{0}$ coupling
regime. And it is worth
noting that, the coupling in the second-layer environment can stimulate the multiple
transitions from Markovian to non-Markovian dynamics and from no-speedup
evolution to speedup evolution. This results hold independently
on the nature of the second-layer environment. Besides, we also can be surprised to find that,
compared with the absence of the coupling between the modes $m_{1}$ and $m_{2}$,
the introduction of coupling in the second-layer
environment cannot play a beneficial role on the non-Markovian speedup dynamics behavior
of the system. To further illustrate the reasons behind the above
results, we try to give a discussion for this problem based on the
competitive relationship between $\kappa$ and $\Omega$.
To be concrete, when the coupling strength $\Omega$ plays the dominating role in the evolution of system,
the non-Markovian speedup dynamics behavior of the system is inhibited and therefore eventually
leads to Markovian no-speedup dynamics behavior. While $\kappa$ acts mainly, the non-Markovian dynamics behavior of the system can be activated.
So the alternating effects of the coupling strength $\kappa$ and $\Omega$ on the dynamic
behaviors of the system can lead to the increasing and decreasing behaviors of the non-Markovianity.
Furthermore, it also explains how the coupling between the two layers affects the non-Markovian
speedup dynamics behavior depends on the nature of the second-layer environment.
For the memoryless nature of the second-layer environment,
the non-Markovianity and the capacity for potential speedup of the system become
greater as increasing $\kappa$. Differently, when the second-layer environment has a memory-keeping nature,
the non-Markovianity and the capacity for potential speedup of the system can be improved and weakened
as increasing the coupling strength $\kappa$ between the two layers.

\section{\textbf{{Acknowledgements}}}
This work was supported by NSFC under grants Nos. 11574022,
11434015, 61227902, 61835013, 11611530676, KZ201610005011,
the National Key R\&D Program of China under grants Nos. 2016YFA0301500,
SPRPCAS under grants No. XDB01020300, XDB21030300.

K. Xu and G. F. Zhang contributed equally to this work.


\begin{thebibliography}{99}

\bibitem{Caruso1}F. Caruso, V. Giovannetti, C. Lupo, and S. Mancini, Rev. Mod.
Phys. \textbf{86}, 1203 (2014).
\bibitem{Lo Franco2}R. Lo Franco, B. Bellomo, S. Maniscalco, and G. Compagno,
Int. J. Mod. Phys. B \textbf{27}, 1345053 (2013).
\bibitem{Rivas3} A. Rivas, S. F. Huelga, and M. B. Plenio, Rep. Prog. Phys. \textbf{77}, 094001 (2014).
\bibitem{Lee4} H. Lee, Y. C. Cheng, and G. R. Fleming, Science \textbf{316}, 1462 (2007).
\bibitem{Cederbaum5} L. S. Cederbaum, E. Gindensperger, and I. Burghardt, Phys.
Rev. Lett. \textbf{94}, 113003 (2005).
\bibitem{Petruccione}H. P. Breuer and F. Petruccione, \textit{Theory of Open Quantum Systems} (Oxford University Press, New York, 2002).
\bibitem{Vega6}I. de Vega and D. Alonso, Rev. Mod. Phys. \textbf{89}, 015001 (2017).
\bibitem{Vacchini7} H. P. Breuer, E. M. Laine, J. Piilo, and B. Vacchini, Rev. Mod. Phys. \textbf{88}, 021002 (2016).
\bibitem{Lorenzo}S. Lorenzo, F. Plastina, and M. Paternostro, Phys. Rev. A \textbf{88}, 020102(R) (2013).
\bibitem{Plenio}A. Rivas, S. F. Huelga, and M. B. Plenio, Phys. Rev. Lett. \textbf{105}, 050403 (2010).
\bibitem{Caruso}F. Caruso, V. Giovannetti, C. Lupo, and  S. Mancini, Rev. Mod. Phys. \textbf{86}, 1203 (2014).
\bibitem{Laine}E. M. Laine, J. Piilo, and H. P. Breuer, Europhys. Lett. \textbf{92}, 60010 (2010).
\bibitem{Dajka}J. Dajka, and J. Luczka, Phys. Rev. A \textbf{82}, 012341 (2010).
\bibitem{Smirne}A. Smirne, H. P. Breuer, J. Piilo, and B. Vacchini, Phys. Rev. A \textbf{82}, 062114 (2010).
\bibitem{Luo}S. Luo, S. Fu, and H. Song, Phys. Rev. A \textbf{86}, 044101 (2012).
\bibitem{Luxm}X. M. Lu, X. Wang, and C. P. Sun, Phys. Rev. A \textbf{82}, 042103 (2010).
\bibitem{Chruscinski2}D. Chruscinski, and S. Maniscalco, Phys. Rev. Lett. \textbf{112}, 120404 (2014); C. Addis, B. Bylicka, D. Chruscinski, and S. Maniscalco, Phys. Rev. A \textbf{90}, 052103 (2014).
\bibitem{SWibmann}S. Wi{\ss}mann, A. Karlsson, E. M. Laine, J. Piilo, and H. P. Breuer, Phys. Rev. A \textbf{86}, 062108 (2012).
\bibitem{Lijg}J. G. Li, J. Zou, and B. Shao, Phys. Rev. A \textbf{81}, 062124 (2010).
\bibitem{Alimm}M. M. Ali, P. Y. Lo, M. W. Y. Tu, and W. M. Zhang, Phys. Rev. A \textbf{92}, 062306 (2015).

\bibitem{Wang0}K. Xu, Y. J. Zhang, Y. J. Xia, Z. D. Wang, and H. Fan, Phys. Rev. A \textbf{98}, 022114 (2018).



\bibitem{Smirne0}A. Smirne, D. Brivio, S. Cialdi, B. Vacchini, and M. G. A. Paris, Phys. Rev. A \textbf{84}, 032112 (2011).
\bibitem{Franco00}R. Lo Franco, B. Bellomo, E. Andersson, and G. Compagno, Phys. Rev. A \textbf{85}, 032318 (2012).
\bibitem{Laine2}E. M. Laine, H. P. Breuer, J. Piilo, C. F. Li, and G. C. Guo, Phys. Rev. Lett. \textbf{108}, 210402 (2012).
\bibitem{Liu4}B. H. Liu, D. Y. Cao, Y. F. Huang, C. F. Li, G. C. Guo, E. M. Laine, H. P. Breuer, and J. Piilo, Sci. Rep. \textbf{3}, 1781 (2013).
\bibitem{Man1}Z. X. Man, Y. J. Xia, and R. Lo Franco, Phys. Rev. A \textbf{92}, 012315 (2015).
\bibitem{Man2}Z. X. Man, N. B. An, and  Y. J. Xia, Phys. Rev. A \textbf{90}, 062104 (2014).
\bibitem{Zou}Y. J. Zhang, X. B. Zou, Y. J. Xia, G. C. Guo, Phys. Rev. A, \textbf{82}, 022108 (2010).
\bibitem{Addis}C. Addis, G. Brebner, P. Haikka and S. Maniscalco, Phys. Rev. A \textbf{89}, 024101 (2014).
\bibitem{Xue}S. B. Xue, R. B. Wu, W. M. Zhang, J. Zhang, C. W. Li, and T. J. Tarn, Phys. Rev. A \textbf{86}, 052304 (2012).
\bibitem{D'Arrigo9}A. D'Arrigo, R. Lo. Franco, G. Benenti, E. Paladino, and G. Falci, Ann. Phys. \textbf{350}, 211 (2014).
\bibitem{Bylicka}B. Bylicka, D. Chruscinski, and S. Maniscalco, Sci. Rep. \textbf{4}, 5720 (2014).
\bibitem{Xiang}Z. L. Xiang, S. Ashhab, J. You, and F. Nori, Rev. Mod. Phys. \textbf{85}, 623 (2013).
\bibitem{Bennett}C. H. Bennett, and D. P. Divincenzo, Nature \textbf{404}, 247-255 (2000).
\bibitem{Xu11}J. S. Xu, K. Sun, C. F. Li, X. Y. Xu, G. C. Guo, E. Andersson, R. Lo Franco, and G. Compagno, Nat. Commun. \textbf{4}, 2851 (2013).
\bibitem{Aaronson}B. Aaronson, R. Lo Franco, and G. Adesso, Phys. Rev. A \textbf{88}, 012120 (2013).
\bibitem{Duan}L. M. Duan, M. D. Lukin, J. I. Cirac, and P. Zoller, Nature \textbf{414}, 413-418 (2001).
\bibitem{sunfangwen}Y. Dong, Y. Zheng, S. Li, C. C. Li, X. D. Chen, G. C. Guo, and F. W. Sun,
njp Quantum information \textbf{4}, 3 (2018).
\bibitem{Piilo0} H. P. Breuer, E. M. Laine, and J. Piilo, Phys. Rev. Lett. \textbf{103}, 210401 (2009).
\bibitem{Plastina} S. Lorenzo, F. Plastina, and M. Paternostro, Phys. Rev. A \textbf{88}, 020102 (2013).
\bibitem{Huelga0} $\acute{A}$. Rivas, S. F. Huelga, and M. B. Plenio, Phys. Rev. Lett. \textbf{105}, 050403 (2010).
\bibitem{Laine0} E. M. Laine, J. Piilo, and H. P. Breuer, Phys. Rev. A \textbf{81}, 062115 (2010).
\bibitem{Chru?ci¨½ski0} D. Chru$\acute{s}$ci$\acute{n}$ski, A. Kossakowski, and $\acute{A}$. Rivas,
Phys. Rev. A \textbf{83}, 052128 (2011).
\bibitem{Maniscalco0} D. Chru$\acute{s}$ci$\acute{n}$ski and S. Maniscalco, Phys. Rev. Lett. \textbf{112}, 120404 (2014).
\bibitem{Hall0} M. J. W. Hall, J. D. Cresser, L. Li, and E. Andersson, Phys. Rev. A \textbf{89}, 042120 (2014).
\bibitem{Chiuri0} A. Chiuri, C. Greganti, L. Mazzola, M. Paternostro, and P. Mataloni, Sci. Rep.\textbf{2}, 968 (2012).
\bibitem{Huang} B. H. Liu, L. Li, Y. F. Huang, C. F. Li, G. C. Guo, E. M. Laine, H. P. Breuer, and J. Piilo, Nat. Phys. \textbf{7}, 931 (2011).
\bibitem{Bernardes} N. K. Bernardes, J. P. S. Peterson, R. S. Sarthour, A. M. Souza,
C. H. Monken, I. Roditi, I. S. Oliveira, and M. F. Santos, Sci. Rep. \textbf{6}, 33945 (2016).

\bibitem{Fan}Y. J. Zhang, Y. J. Xia, and H. Fan, Europhys. Lett. \textbf{116}, 30001 (2016).
\bibitem{Sun}Z. Sun, J. Liu, J. Ma, and X. Wang, Sci. Rep. \textbf{5}, 8444 (2015).
\bibitem{Meng}X. Meng, C. Wu, and H. Guo, Sci. Rep. \textbf{5}, 16357 (2015).
\bibitem{Zhangyj3}Y. J. Zhang, W. Han, Y. J. Xia, J. P. Cao, and H. Fan, Sci. Rep. \textbf{4}, 4890 (2014).
\bibitem{Xuzy}Z. Y. Xu, S. Luo, W. L. Yang, C. Liu, and S. Q. Zhu, Phys. Rev. A \textbf{89}, 012307 (2014).
\bibitem{Anjh2}H. B. Liu, W. L. Yang, J. H. An, and Z. Y. Xu, Phys. Rev. A \textbf{93}, 020105R (2016).
\bibitem{Zhangyj4}Y. J. Zhang, W. Han, Y. J. Xia, J. P. Cao, and H. Fan, Phys. Rev. A \textbf{91}, 032112 (2015).
\bibitem{Xuzy2}Z. Y. Xu, and S. Q. Zhu, Chinese Physics Letters \textbf{31}, 2 (2014).
\bibitem{Cianciaruso}M. Cianciaruso, S. Maniscalco, and G. Adesso, Phys. Rev. A \textbf{96}, 012105 (2017).
\bibitem{Deffner}S. Deffner and E. Lutz, Phys. Rev. Lett. \textbf{111}, 010402 (2013).
\bibitem{Cimmarusti}A. D. Cimmarusti, Z. Yan, B. D. Patterson, L. P. Corcos, L. A. Orozco, and S. Deffner, Phys. Rev. Lett. \textbf{114}, 233602 (2015).
\bibitem{Hanson1}R. Hanson, V. V. Dobrovitski, A. E. Feiguin, O. Gywat, and D. D. Awschalom, Science \textbf{320}, 352 (2008).
\bibitem{Hanson2}R. Hanson, L. P. Kouwehoven, J. R. Petta, S. Tarucha, and L. M. K. Vandersypen, Rev. Mod. Phys. \textbf{79}, 1217 (2007).
\bibitem{Pla}J. J. Pla,  K. Y. Tan, J. P. Dehollain, W. H. Lim, J. J. L. Morton, D. N. Jamieson, A. S. Dzurak, and A. Moreello, Nature (London) \textbf{489}, 541 (2012).
\bibitem{Chekhovich}E. A. Chekhovich,  M. N. Makhonin, A. I. Tartakovskii, A. Yacoby, H. Bluhm, K. C. Nowack and L. M. K. Vandersypen, Nat. Mater. \textbf{12}, 494 (2013).
\bibitem{Tyryshkin}A. M. Tyryshkin,  S. Tojo, J. J. L. Morton, H. Riemann, N. V. Abrosimov, P. Becker, H. J. Pohl, T. Schenkel, M. L. W. Thewalt, K. M. Itoh, and S. A. Lyon, Nat. Mater. \textbf{11}, 143 (2012).
\bibitem{Mat}T. T. Ma, Y. S. Chen, T. Chen, S. R. Hedemann, and T. Yu, Phys. Rev. A \textbf{90}, 042108 (2014).
\bibitem{Nguyen}Z. X. Man, N. B. An, and Y. J. Xia, Opt. Express \textbf{23}, 5763 (2015).





\bibitem{8}M. M. Taddei, B. M. Escher, L. Davidovich, and R. L. de Matos Filho, Phys. Rev. Lett. \textbf{110}, 050402 (2013).
\bibitem{08}A. del Campo, I. L. Egusquiza, M. B. Plenio, and S. F. Huelga, Phys. Rev. Lett. \textbf{110}, 050403 (2013).



\end{thebibliography}
\end{document}